\documentclass[
    twocolumn,
    showpacs,preprintnumbers,amsmath,amssymb]{revtex4}

\usepackage{graphicx,subfigure}
\usepackage{dcolumn}
\usepackage{bm}
\usepackage{hyperref}
    
\usepackage{textcomp}
\usepackage{nicefrac}

\usepackage{mathptmx}
\usepackage{scrtime}
\usepackage{epstopdf}
\usepackage[normalem]{ulem} 
\usepackage{color}
\usepackage[color=red!40,linecolor=red,textsize=scriptsize]{todonotes}

\newcommand{\mT}{\, \mathrm{mT}}                    
\newcommand{\nm}{\, \mathrm{nm}}                   

\newcommand{\GHz}{\, \mathrm{GHz}}

\newcommand{\mum}{\, \mbox{\textmu}\mathrm{m}}

\newcommand{\MS}{\ensuremath{M_\mathrm{S}}}                     
\newcommand{\muMS}{\ensuremath{\mu_0 M_\mathrm{S}}}     
\newcommand{\Hext}{\ensuremath{H_\mathrm{ext}}}                     
\newcommand{\muHext}{\ensuremath{\mu_0 H_\mathrm{ext}}} 

\begin{document}

\title{Spin-wave modes and band structure of rectangular CoFeB antidot lattices}

\author{Benjamin Lenk}
\email{blenk@gwdg.de}
\author{Nils Abeling}
\author{Jelena Panke}
\author{Markus M\"{u}nzenberg}
\affiliation{I. Institute of Physics, University of G\"{o}ttingen, Friedrich-Hund-Platz 1, 37077~G\"{o}ttingen, Germany}


\begin{abstract}
We present an investigation of rectangular antidot lattices in a CoFeB film. Magnonic band structures are numerically calculated, and band gaps are predicted which shift in frequency by $0.9\GHz$ when rotating the external field from the long to the short axis of the unit cell. We demonstrate by time-resolved experiments that magnonic dipolar surface modes are split in frequency by $0.6\GHz$ which agrees well with the theoretical prediction. These findings provide the basis for directional spin-wave filtering with magnonic devices.
\end{abstract}



\maketitle

The use of spin waves opens up routes to new computing devices with advantages over today's CMOS-based technology. Although magnetic damping is comparably high (requiring small-scale devices), Joule's heating caused by electron currents is avoided~\cite{Khitun2010}.
Much research has been devoted to interferometer-like structures~\cite{Schneider08}, inspired by the possibility of (local) spin-wave phase manipulation by Oersted fields~\cite{Kostylev2005}.
On the other hand, effective spin-wave filters can be designed on the basis of ferromagnetic stripes with modulated widths~\cite{Lee2009:mod:Py:stripes, ChumakAPL2009, Ciubotaru2012}.
One concept of spin-wave excitation and detection is based on rf-antennas which, however, cannot be put in arbitrary proximity, due to inductive coupling. If, instead, the excitation was achieved by intense light pulses as implemented in heat-assisted recording in modern hard disc drives~\cite{Kryder2008, Stipe2010}, only one antenna would be needed for detection.
Hence, effective mechanisms for spin-wave selection from the broad-band (laser) excitation are required.
It has been shown that magnonic crystals inhibit the necessary features: two-dimensional antidot lattices show Bloch-like modes with distinct wave vector, which is in turn tunable by the magnonic lattice parameter~\cite{Ulrichs2010}. These modes propagate in the Damon-Eshbach (DE) geometry, i.e.\ with the wave vector~$k_\mathrm{DE}$ perpendicular to the external magnetic field~\Hext~\cite{de61}.
They have for example been used for spin-wave imaging~\cite{Mansfeld2012}.

A transition between different magnonic crystals may allow the scattering of one magnonic mode into another.
In particular, when propagation takes place across a (one-dimensional) interface, spin-wave tuning or filtering is viable, if the magnonic lattices in question are of similar character.
This can be achieved in rectangular lattices, i.e., if the orthogonal unit vectors of the antidot lattice~$a_1$ and $a_2$ differ in length. In such magnonic materials we demonstrate experimentally how the lattice anisotropy can be employed to change the spin-wave characteristics.
A rotation of the magnetic field from along the long axis of the rectangular lattice to the short axis allows to decrease the spin-wave frequency by $\approx 0.6\GHz$ rerouting the spin wave by $90^\circ$.
Namely, we observe spin-wave splitting at the Brillouin zone boundary which opens routes to magnonic spin-wave filter devices tunable by rotating the magnetic field.
\begin{figure}%
\centering%
\includegraphics[width=3.3in]{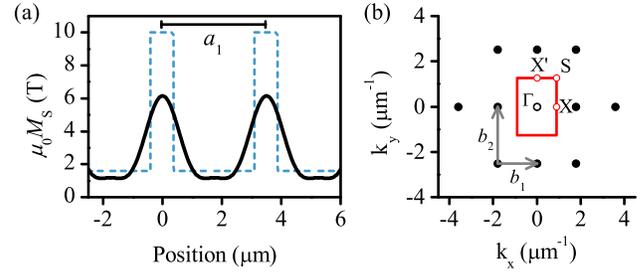}%
\caption{Uniform mode analysis in rectangular antidot lattices. (a)~Calculated magnetization profile using \hyperref[eq:ms-fourier]{Eq.~(\ref*{eq:ms-fourier})} along a high-symmetry direction (solid black line). The lattice parameter~$a_1 = 3.5\mum$ is depicted as well as the idealized profile (blue dashed line). In (b), the respective reciprocal lattice vectors in Fourier space are plotted (black points). The Brillouin zone boundary is given by the solid red line, high symmetry points~$\Gamma$, X, S, and X' are marked in white. Reciprocal lattice unit vectors (gray arrows) are $b_1 = 2\pi / a_1$ and $b_2 = 2\pi / a_2$, respectively.}%
\label{fig:reciprocal-and-profile}%
\end{figure}%

In order to develop a theoretical understanding, band structure calculations are performed, the numerical formalism of which has been presented in detail in Ref.~\cite{Lenk2011}. In brief, the Landau-Lifshitz-Gilbert (LLG) equation of motion is solved by a plane-wave method as developed by Puszkarski and co-workers~\cite{Vasseur1996, Krawczyk08}.
For the case of a thin ferromagnetic film, dynamic magnetic modes can be --~under neglection of the exchange interaction~-- assumed to be uniform across the film thickness~\cite{Hurben1995}.
In the lateral direction, the periodic modulation of the sample's magnetization between film and antidots is achieved by a Fourier synthesis~\cite{Vasseur1996}
\begin{equation}\label{eq:ms-fourier}
    M_\mathrm{S}( \boldsymbol{r} )=\sum_{\boldsymbol{G}} M_\mathrm{S}( \boldsymbol{G} ) e^{i \boldsymbol{G r} },
\end{equation}
where $\boldsymbol{G}$ is a two-dimensional vector of the reciprocal lattice.
The profile of $M_\mathrm{S}$ which has been used in the calculations is plotted as a black line in \hyperref[fig:reciprocal-and-profile]{Fig.~\ref*{fig:reciprocal-and-profile}(a)}. The constituting reciprocal lattice vectors are shown in \hyperref[fig:reciprocal-and-profile]{Fig.~\ref*{fig:reciprocal-and-profile}(b)}.
These provide a compromise between a desirably well approximation of the stepwise magnetization profile on the one hand, yet fulfilling the initial assumption of mode uniformity ($\lambda \gg t$, with spin-wave length~$\lambda$ and thickness~$t$) on the other hand.
\begin{figure*}%
\centering%
\includegraphics[width=6.3in]{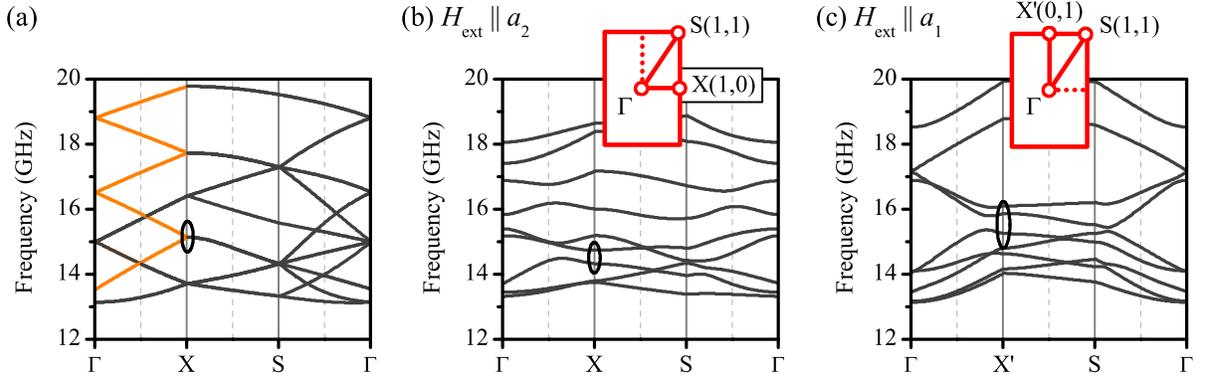}
\caption{Numerically calculated band structures for rectangular magnonic crystals in a CoFeB matrix. In (a), the continuous film is presented with a vanishing antidot radius~$R=0$, maintaining the periodicities $a_1 = 3.5\mum$ and $a_2 = 2.5\mum$. The external field is applied parallel to $a_2$ and a folding of the Damon-Eshbach dispersion into the first Brillouin zone is observed (orange line). Plots in~(b) and~(c) show the band structures for a non-vanishing radius~$R=400\nm$ and \Hext\ applied along $a_2$ and $a_1$, respectively. The insets illustrate the geometry, where the dotted red lines depict the direction of \Hext. For spin-wave propagation perpendicular to \Hext\ and \MS, i.e. in the Damon-Eshbach geometry, band gaps open up at the X- and X'-points (black ellipses). These shift in frequency by $0.9\GHz$.}%
\label{fig:rectangles:band-calc}%
\end{figure*}%

Other parameters in the calculations were $\muMS = 1.6\,\mathrm{T}$, $\muHext =130\mT$, and $g=2.04$.
An antidot lattice with $a_1 = 3.5\mum$ and $a_2 = 2.5\mum$ was the basis for both the calculations as well as the experiments to be discussed later in this manuscript.
Plotted in \autoref{fig:rectangles:band-calc} are the numerically obtained results of the band structure calculations.
In (a), the band structure of a continuous film is presented which is modeled in the limit $R \to 0$, with unchanged lattice parameters~$a_i$.
The external field was applied parallel to $a_2$ and a folding of the Damon-Eshbach dispersion into the first Brillouin zone is found as expected from solid state theory (orange line).

A finite antidot radius of $R=400\nm$ leads to the magnonic band structures shown in \hyperref[fig:rectangles:band-calc]{Fig.~\ref*{fig:rectangles:band-calc}(b)} and \hyperref[fig:rectangles:band-calc]{(c)}.
For the two plots the external field was applied along either of the unit vector directions of the antidot lattice. The paths to the X- and X'-points in reciprocal space hence correspond to DE surface waves propagating perpendicular to \Hext\ and \MS.
When approaching the Brillouin zone boundary, the bands for these modes flatten out and culminate in band gaps at~X and X' as marked by the black ellipses in the graph.
By means of the aspect ratio of the rectangular lattice vectors, also the energy (i.e.\ frequency) of the Bloch-like modes at X and X' can be shifted: given the condition $k_\mathrm{DE} = \pi /a_i$ the DE-frequency will change when changing $a_i$, i.e.,\ when propagation of the DE modes is along either of the two lattice unit vectors.
Namely, a frequency shift of $0.9\,\mathrm{GHz}$ is calculated.

\begin{figure*}%
\centering%
\includegraphics[width=6in]{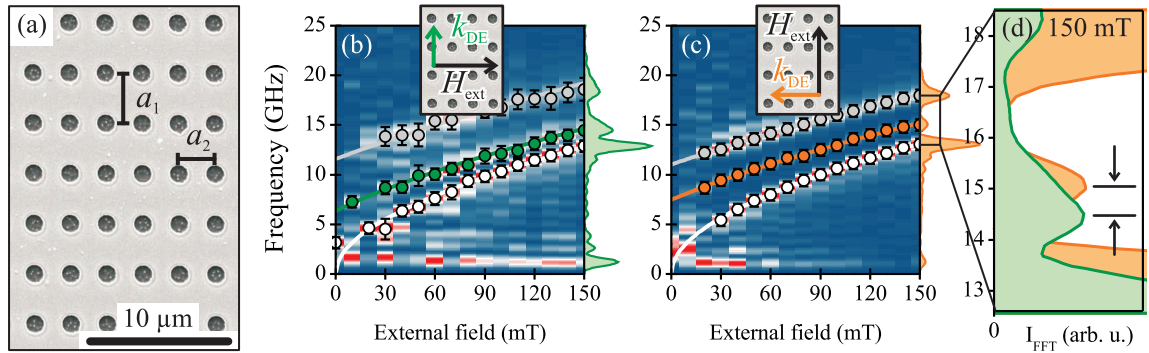}%
\caption{Experiments on magnonic spin-wave modes in rectangular antidot lattices. In (a), an SEM image of the structured CoFeB film with $a_1=3.5\mum$, $a_2=2.5\mum$, and $R=400\nm$ is given. The plots (b) and (c) show the Fourier power of the magnetic precession observed after optical excitation with fs-laser pulses. The peak positions have been determined and are given by the points (white: uniform precession; gray: PSSW; green and orange: magnonic Damon-Eshbach mode). The fitted dispersion curves are represented by solid lines~\cite{Lenk2010}. Detailed in~(d) is the frequency shift of the magnonic Damon-Eshbach mode at an applied field of $\muHext=150\mT$ between the cases $\Hext \perp a_1$ and $\Hext \perp a_2$. It accounts to $0.55(4)\GHz$ and agrees reasonably well with the simulations from \autoref{fig:rectangles:band-calc}.}%
\label{fig:rectangles:trmoke}%
\end{figure*}

In the following we will describe an experimental evaluation of the numerical results using femtosecond laser pulses.
An all-optical approach was utilized, where one can make use of the very broad band, neither frequency- nor $k$-selective excitation of spin waves~\cite{JakobJPD2008}. In principle, such an experiment will show those spin-wave modes with the highest density of states~(DOS).
According to general solid state theory, the flattened bands found above should lead to an increased DOS and should therefore resemble the rectangular anisotropy. Since the bands shift in \autoref{fig:rectangles:band-calc}, a change of the DOS should be experimentally observed in an altered population of spin-wave modes.
For the experiments, a Co$_{20}$Fe$_{60}$B$_{20}$ film with a thickness of $t=50\nm$ was magnetron-sputtered onto a Si(100) substrate and passivated with $3\nm$ of ruthenium.
With a focussed beam of Ga-ions~(FIB) a rectangular magnonic crystal was created using the same structural parameters $\{a_1, a_2, R\}$ as in the calculations. The overall size of the structured area was $150 \times 150 \mum^2$, considerably larger than the pump and probe laser spot sizes in the experiment ($60\mum$ and $15\mum$, respectively).
In \hyperref[fig:rectangles:trmoke]{Fig.~\ref*{fig:rectangles:trmoke}(a)} an SEM image of the sample is shown.

Data analysis followed a scheme as presented in Ref.~\cite{Lenk2010}. We refrain from plotting reference data on a continuous CoFeB film here. These have already been shown in~\cite{Ulrichs2010} for an identical specimen.
Plotted in \hyperref[fig:rectangles:trmoke]{Fig.~\ref*{fig:rectangles:trmoke}(b)} and \hyperref[fig:rectangles:trmoke]{(c)} are the Fourier-analyzed TRMOKE data as recorded on a rectangular antidot lattice milled into a CoFeB film.
Depicted by the SEM insets is the orientation of the external magnetic field~\Hext\ with respect to the two-dimensional magnonic crystal.

The points in \hyperref[fig:rectangles:trmoke]{Fig.~\ref*{fig:rectangles:trmoke}(b)} and \hyperref[fig:rectangles:trmoke]{(c)} represent the peak positions and are attributed to the uniform ($k=0$) Kittel mode (white), the exchange-dominated perpendicular standing spin waves (PSSW, gray), and the magnonic Bloch-like DE-modes exclusively excited on periodically structured samples (green: $\Hext\parallel a_2$, and orange: $\Hext\parallel a_1$).
We would like to emphasize that while PSSW and Kittel modes are the only ones observed on a continuous film, the DE-modes originate from the magnonic crystal's periodicity in the propagation direction (perpendicular to \Hext\ and \MS)~\cite{Ulrichs2010}.
Also included as correspondingly colored solid lines are the theoretically expected dispersions~\cite{Lenk2010, Lenk2011}. For simplicity, here we only state the dipolar Damon-Eshbach dispersion which reads~\cite{de61}
\begin{equation}\label{eq:rectangles:de}
\left(\frac{2\pi f_\mathrm{DE}}{\gamma\mu_0}\right)^2= H_x \left(H_x + M_\mathrm{S} \right)+\frac{M_\mathrm{S}^2}{4}\Big(1-e^{-2 |k_\mathrm{DE}| t}\Big).
\end{equation}
Therein, $t$ is the thickness, $\mu_0M_\mathrm{S} = 1.6\, \mathrm{T}$ is the saturation magnetization as stated above, and $H_x = \Hext \cos \phi$ is the projection of the canted external field onto the film plane ($\phi=30^\circ$ for the experiments presented here). Therefore, as the only free parameter the wave vector~$k_\mathrm{DE}$ remains.

Given by the solid green and orange lines is the fit of the Damon-Eshbach dispersion~(\ref{eq:rectangles:de}) to the experimentally determined dispersion~$f_\mathrm{DE}(\Hext)$.
In both cases of $\Hext \perp a_{i}$ ($i=1,2$) the fits yield the Damon-Eshbach wave vectors~$k_\mathrm{DE,1} = 0.87(7)\mum^{-1} = 0.97\times\pi/ a_1$ and $k_\mathrm{DE,2} = 1.23(7)\mum^{-1} = 0.98\times\pi/ a_2$.
Hence, not only can one single magnonic mode be defined in the structures~\cite{Ulrichs2010}.
Instead, merely changing the relative orientation between external field and antidot lattice by $90^\circ$ is sufficient to excite a different magnonic spin-wave mode.
This is accompanied by a frequency shift further detailed in \hyperref[fig:rectangles:trmoke]{Fig.~\ref*{fig:rectangles:trmoke}(d)} which contains the Fourier spectra of the TRMOKE measurements performed at $\Hext=150\mT$.
The shift of the magnonic mode's frequency is marked by the black arrows and accounts to $0.55\pm0.04\GHz$, which is similar to the value expected from the calculations in \autoref{fig:rectangles:band-calc}.


The bosonic character of spin waves becomes apparent in the condensation-like excitation in the TRMOKE experiment~\cite{marija07}. As a consequence, selected spin-wave excitation is possible and processing schemes which employ the spin-wave propagation for manipulation purposes on top of mere transport can be applied~\cite{Schneider08, Khitun2010}.
In view of the results presented in this manuscript, the interplay between the intrinsic anisotropy of the dipolar modes' dispersion $\omega |_{k\perp M} \neq \omega |_{k\parallel M}$ and the (rectangular) anisotropy stemming from the magnonic crystals can further be employed.
Namely, frequency splitting of spin waves becomes feasible, with the direction of the applied magnetic field as the external control parameter.
By means of the magnetic field, the propagation direction of the spin waves is changed, accompanied by the frequency shift described above.

In conclusion, we expect from numerical calculations the opening of magnonic band gaps in the Damon-Eshbach geometry and verify this with TRMOKE results that show the optical excitation of dynamic modes with wave vectors at the Brillouin zone boundary.
Thus, a controlled excitation of selected spin waves can be achieved by rotation of the external field.
In a more farsighted view, interfaces between respective magnonic crystals provide interesting perspectives: a reflection of spin waves may be observed due to an abrupt change of the magnonic index of refraction~\cite{Kampfrath2010PRA,Neusser2011}.
Similarly, the spin-wave splitting observed here hints towards directional switching devices for spin waves defined by rectangular (i.e.\ anisotropic) magnonic crystals.

\end{document}